# An Analysis of Honeypots and their Impact as a Cyber Deception Tactic

Daniel Zielinski, Hisham A. Kholidy
State University of New York (SUNY) Polytechnic Institute,
College of Engineering, Network and Computer Security Department, Utica, NY USA
zielind@sunypoly.edu, hisham.kholidy@sunypoly.edu

*Abstract—* **This paper explores deploying a cyber honeypot system to learn how cyber defenders can use a honeypot system as a deception mechanism to gather intelligence. Defenders can gather intelligence about an attacker such as the autonomous system that the attacker's IP is allocated from, the way the attacker is trying to penetrate the system, what different types of attacks are being used, the commands the attacker is running once they are inside the honeypot, and what malware the attacker is downloading to the deployed system. We demonstrate an experiment to implement a honeypot system that can lure in attackers and gather all the information mentioned above. The data collected is then thoroughly analyzed and explained to understand all this information. This experiment can be recreated and makes use of many open-source tools to successfully create a honeypot system.**

*Keywords—honeypots, deception mechanism, autonomous system, deceptive honeypot system, open-source tools*

## I. INTRODUCTION

Cyber honeypots can be very useful tools when trying to lure in attackers and gain a defensive advantage by learning how attackers are operating and what information they are seeking. Honeypots are very common systems that many enterprise organizations use to improve their security program and gain insight into the common attacks they face.

When using a honeypot, defenders can gain various amounts of information about the attackers targeting them. Defenders can then use this information to better protect their "real" environments. Defenders can gain insight into the most popular types of attacks that are being used to break into their honeypot systems and then make sure that their actual systems are properly protected against these attacks. Defenders can also learn what data the attackers are trying to reach, or what exactly the motive of the attackers is. In addition, defenders will be able to learn the different types of malware that attackers are installing onto the honeypot systems.

Modern honeypots are not able to look or be configured the same way that they have in the past. This is because over time attackers started learning different ways to detect if what they were attacking is a honeypot. Once an attacker figures out that what they are attacking is a honeypot, they will no longer use their time or resources on the honeypot. This will then make it where the defender will not be able to gain the information they are seeking about the attacker, and the attacker will not use up a great number of resources on the honeypot. Therefore, modern honeypot systems must be deceptive. They need to look, run, and operate like the "real" system. An effective way of doing so is by looking at your actual system and then placing data that looks the same but is fake into your honeypot. This will make attackers spend a lot of time trying to exploit your honeypot and make them spend time searching for artifacts such as valid credentials. This will also help dry up a lot of the attackers' resources and make them waste a lot of money. By the time the attackers realize that they have been caught in a honeypot trap, they may become extremely frustrated and decide to use their remaining resources on a different target to prevent them from falling into another honeytrap that you or your organization may have set up.

In the next sections, We will be explaining the details of a honeypot and the architecture of a modern deceptive honeypot system. We will also go into the different types of information a defender can gain from using a honeypot and how this information can be leveraged. We then conduct an experiment in which we use a Virtual Private Server to deploy an open-source honeypot system that utilizes many different honeypot containers to create an effective system. This system also utilizes many other features such as an IPS/IDS and a data visualization dashboard to make it easier to analyze data.

## II. BACKGROUND

In this section, we will discuss the network architecture of where to place a honeypot, different types of honeypots, types of data honeypots can collect, and techniques to achieve deception within the honeypot itself.

*A. Honeypot Network Architecture*

When deciding where to place a honeypot on a network you must be very careful. Placing a honeypot inside an internal network can be a major security design flaw. This is because it would allow an attacker to easily gain access to your internal network by exploiting the honeypot. An attacker would then be able to move laterally throughout the network to find real systems and servers. In essence, this would make it much easier for the attacker because you would be "inviting" them into your internal local area network (LAN). Instead, a honeypot should be placed inside a network's Demilitarized Zone (DMZ).

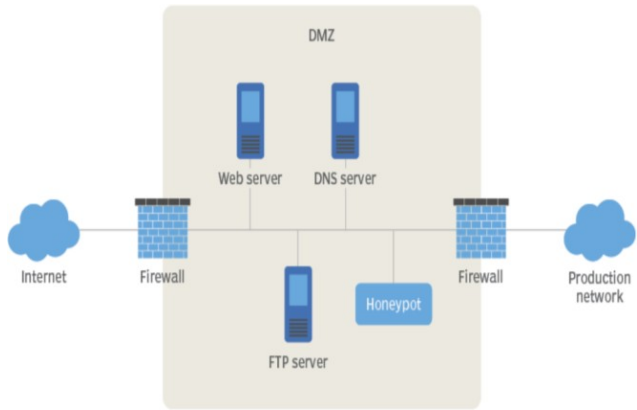

Figure 1: Architecture of a network with a honeypot deployed [1].

A DMZ is an isolated part of the network that is connected to the internet and is typically where public-facing services are placed. In this example, a web server, DNS server, and FTP server are all placed inside the DMZ along with the honeypot [1]. The DMZ is separated on both sides with a firewall. The firewall between the DMZ and Internet is typically a weaker firewall that allows traffic to outside users to do tasks like visiting a website hosted or interacting with available files. This firewall will not allow in all traffic but will have the necessary ports open to allow functionality. The firewall between the DMZ and the production network is a much more hardened firewall. This firewall will not let outside, or unauthorized users bypass it easily. In an enterprise setting, most organizations will have an Intrusion Detection System (IDS), or Intrusion Prevention System (IPS) configured to alert on attacks being made against this firewall [2]. This is to keep cyber analysts alert of any attacks that might be going on against their network. The production network may have sensitive data flowing through it and may have systems that store important data, so it is important to keep the honeypot isolated from it.

*B. Honeypot Characteristics*

There are different types of modern honeypots that one can deploy into their environments. However, all the honeypots do have some things in common such as being low-maintenance, low cost, and easy to deploy [3]. The reason honeypots need to need to be low maintenance is that, from an organizational perspective, engineers cannot be using up large amounts of their time manually making changes and fixes to the honeypot. They have many other duties to take care of and should be able to look at and analyze data collected from the honeypot easily. This is also connected to why a honeypot is easy to deploy. You should take your time deploying a honeypot to ensure that everything is working properly, but honeypots are easy to deploy because more time and focus should be spent on securing the real environment. If anything goes wrong with your honeypot you should be able to just reboot it and most of the time it will fix itself. Furthermore, a honeypot is often relatively low cost compared to the number of resources an individual or organization may have. Now, one can make a massive honeypot that is extremely expensive, but it would not make any sense from a financial perspective or provide a great amount of additional benefit. Most individuals and organizations deploy honeypots that are not very expensive since they can provide great benefits at a low cost. This also then allows for a great number of financial resources to be spent on other things like employees, software, and other systems.

*C. Types of Honeypots Based on their Design*

Many different types of Honeypots can be deployed with various complexities.

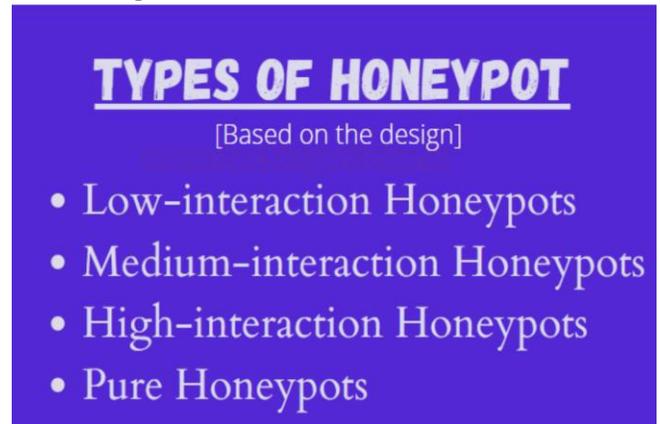

Figure 2: Types of Honeypots [Based on Design] [4].

A pure honeypot is meant to be a full-scale replica of a production environment that contains fake data that is meant to pose as real data. A high-interactive honeypot is like a pure honeypot because it runs many different real-like services, but it does not hold as much data and is not a replica of a full-scale production environment. A mild-interaction honeypot is different than a high-interaction honeypot because it just emulates aspects of the application layer but does not have its own OS. This can be used to stall or confuse attackers who are trying to attack your systems [5]. Finally, a low-interaction honeypot is very lightweight and can be used to match a small number of services and applications that are in use. This type of honeypot can be used to keep track of UDP, TCP, and ICMP ports and services. In which we make can make use of things like fake databases, data, and files as bait to trap attackers to understand the attacks that happen in real-time [4].

*D. Types of Honeypots Based on their Technologies*

There are different types of honeypots based on the deception technologies that they utilize.

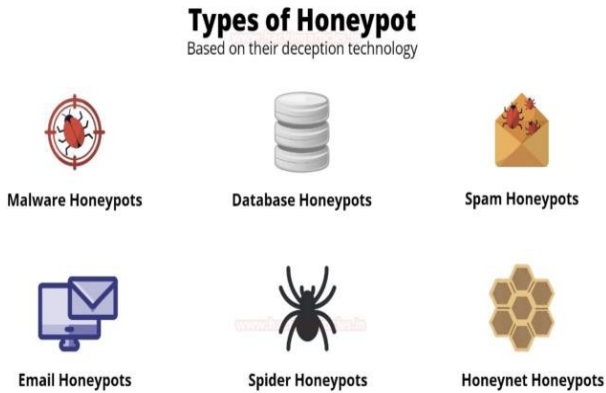

Figure 3: Types of Honeypots [Based on their deception technology] [4].

A malware honeypot is a type of honeypot that is meant to identify and trap malware inside a network or system. A sophisticated malware honeypot will recognize the malware's signature and alert based on its Common Vulnerabilities and Exposures (CVE) ID. It will also keep a count of the CVEs to help recognize the most used malware that is being installed onto a system. A database honeypot is exactly what it sounds like – a honeypot that appears to be a vulnerable database. Often it will be something like an SQL database that will allow things like injection attacks to occur to attract attackers looking to gain sensitive information that can be stored in a database such as credit card numbers. A spider honeypot is installed to trap web crawlers that target web applications with the intent of stealing data. An email honeypot is a fake email server that utilizes hoax email addresses and emails to attract attackers to interact with it. Any suspicious email received by attackers can be scanned, and their email addresses can be then blacklisted on the real email servers. A spam honeypot is like an email honeypot but is meant to attract spammers to exploit vulnerable email elements and give details about their activities. Finally, a honeynet is a honeypot system that can contain various types of honeypots mentioned. It will often aggregate data from all the honeypots into a central location to make alerting and monitoring easier [4].

*E. Data that Honeypots Collect*

Honeypots can collect a large amount of data that can be very useful in establishing a more secure security program as well as serve to be very useful for research purposes. It is often useful to have a service configured that aggregates this data into a dashboard that makes it easier to look at and understand.

One of the most basic artifacts that a honeypot can collect is the attacker's IP address. This is a useful artifact because from the IP address we can see things like the general location of the attacker to help identify if potentially many attacks are coming from a specific area. Additionally, we can see if this attacker is launching many attacks on our systems, and we can see how long they have been attacking our systems [6]. We can also block this IP address from accessing our real network. Finally, from the IP address, we can find the attacker's Autonomous System Number (ASN) to see what Autonomous System the attacker's IP is allocated from.

Another artifact we can gather about the attacker is what Operating System (OS) they are using on their host. We may not be able to figure out the exact version of what OS they are using, but we will be able to figure out a general version of the OS they are using. For instance, we can see if the attacker is running Windows 7 or if the attacker is running a Linux version from 2.2.x-3.x.

Many attackers utilize automation tools to try to break into a server.

Figure 4: A sample Dictionary Attack trying common passwords on an SSH server [7].

They often run dictionary attacks, a type of attack that tries to log in to a server using a list of commonly used or default usernames and passwords, to crack into your system. We can use a honeypot to see what passwords and usernames the attackers are trying to use to log in and even take it a step further by making a count of each username and password attempted to find the most popular credentials attackers are trying to use.

Furthermore, when an attacker gains access to a honeypot system we can see what commands they are running once they break in. This can help us figure out what their motives may be and what they are looking for. We can also see what they are downloading onto the system. They can download worms or viruses onto the system, or maybe even agents that will try to clean up log files to keep them undetected so they can persist more. They even might install things like rootkits to help with this persistence. In essence, any time the attacker attempts to interact with the honeypot system, you can access and track that information.

*F. Honeypot Deception Techniques*

Many different techniques can be utilized to achieve deception within a honeypot system to trick attackers. To start, a commonly used technique is to not allow open access to a server. If attackers can connect to your server on the first try, it is often a sign that you are inviting them in, and it may cause them to be very suspicious. To solve this, make it that after many failed attempts they will finally be able to gain access and login into your system. This will make them think that their brute-forcing or dictionary attack worked, but it still took a while for it to be a success. This will cause them to have less suspicion.

Another more complex technique is to utilize honey credentials.

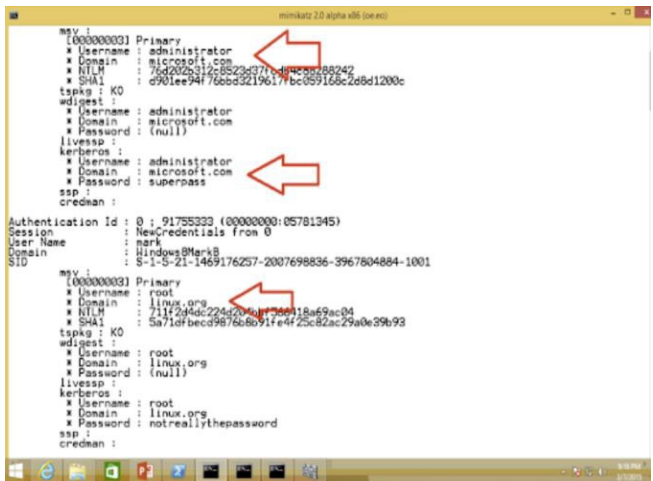
Figure 5: Honey Credentials being stored in memory [8].

Honey credentials help catch malicious actors by injecting fake credentials into a system's memory. When an attacker gains access to your network and finds the honey credentials, they will attempt to use them. Since these credentials don't exist, any attempt to use them can trigger an alert and notify you immediately. In a targeted attack, the attacker will be able to dump recovered honey credentials from the system's memory through privilege escalation or a system flaw. The attacker will then attempt to perform lateral movement into the fake objects, resulting in their exposure and making it easy to trace them [9].

Finally, the most important deception technique is to make the honeypot blend in and look realistic. Earlier we mentioned the different types of honeypots based on their deception technology. It is important to be able to look at these honeypots from the attacker's perspective. For instance, an SSH server must look like an actual SSH server that you or your organization would utilize. If it looks fake no one will take the bait and they will be able to easily detect what they are attacking is a honeypot. You should be configuring and storing data on your honeypot that you would store on your actual systems but just make sure it is fake information [10]. You can install real services you would normally use onto your honeypot, but just make sure they do not contain sensitive information. By having your honeypot blend in, it will keep your attacker occupied and cause them to reveal more information about themselves and what their motives are.

### III. T-POT HONEYPOT SYSTEM FRAMEWORK

In this section, we detail the honeypot system we will be deploying to gather data and intelligence about attackers.

*A. System Fundamentals*

T-Pot is an open-source all-in-one honeypot platform that runs on Debian Linux. The honeypot daemons as well as other support components are dockered. This allows T-Pot to run multiple honeypot daemons and tools on the same network interface while maintaining a small footprint and while constraining each honeypot within its own environment [11]. Documentation to the platform can be found at https://github.com/telekom-security/tpotce. T-Pot uses docker images for 25 different honeypots to create a massive honeypot system. All the honeypots each represent different systems. For instance, cowrie is a medium-to-high interaction SSH and Telnet honeypot designed to log brute force attacks and the shell interaction performed by the attacker [12]. In addition, another example is the honeytrap honeypot is a network security tool written to observe attacks against TCP or UDP services [13]. You can read about all 25 different honeypots deployed in the system on the documentation page for T-Pot. For the most part, understanding each honeypot will not be very important because we will mainly focus on the aggregated data collected for the whole system.

*B. Tools Utilized by T-Pot*

T-Pot uses a variety of different tools to aggregate, alert, monitor, and analyze data collected.

- Cockpit for a lightweight, webui for docker, os, real-time performance monitoring and web terminal.
- Cyberchef a web app for encryption, encoding, compression and data analysis.
- ELK stack to beautifully visualize all the events captured by T-Pot.
- Elasticsearch Head a web front end for browsing and interacting with an Elastic Search cluster.
- Fatt a pyshark based script for extracting network metadata and fingerprints from pcap files and live network traffic.
- Spiderfoot a open source intelligence automation tool.
- Suricata a Network Security Monitoring engine.

Figure 6: Description of tools used by T-Pot [11].

All the tools mentioned in Figure 6 can be further analyzed by examining the information found on the T-Pot documentation page. For our experiment, we will focus on the ELK stack and Suricata tools. ELK stack has a tool named Kibana, which we will utilize to visualize all the data collected by the different honeypots. We can use Kibana to analyze the data for each individual honeypot as well as look at an aggregated view of information collected from all the honeypots deployed in one dashboard [14]. Suricata is a network security monitoring engine that we can use to monitor and alert us about suspicious activity occurring within the system [15]. The information from Suricata will be viewable and aggregated into the Kibana Dashboard. As mentioned previously and displayed in Figure 6 there are many other tools pre-built into T-Pot, but we will not be using them as they are not necessary to understand and analyze the data collected.

*C. Deploying the T-Pot Honeypot System*

The system requirements for deploying the system are as follows: 8 GB Ram, 128 GB SSD, Network via DHCP, and a non-proxied internet connection. You can download the Pre-built ISO Image (~50 MB), or you can create your own ISO Image that allows you to customize the system to fit your needs better. Since we are doing this for research purposes to see what information we can learn about attackers and not using the system in an enterprise environment, xutilized the Pre-built ISO Image.

To deploy the honeypot system, I used a Virtual Private Server (VPS). I uploaded the ISO file to the VPS provider's website and then began installation. The reason I used a VPS is to avoid large amounts of unwanted traffic coming towards my personal network. I also did not want to reveal my actual public IP address to attackers.

Once the ISO image finished installing in my VM the honeypot system was fully functional and ready to lure in attackers. To reach the T-Pot landing page to gain access to all the tools deployed in the system I had to go into a web browser and enter "https://<my.ip>:64297". I then logged in with the credentials created during installation.

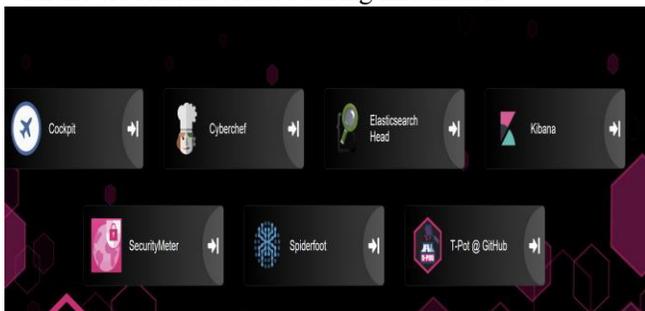

Figure 7: T-Pot Landing Page.

Now that the honeypot is fully deployed attackers will begin to attack the honeypot. We can view data collected from the attacks in the Kibana dashboard. I will wait approximately 3 weeks before analyzing the data just to let a good amount of data accumulate. In the next section, we will test some of the functionality on the honeypot by running some attacks to make sure the honeypot is configured properly before waiting the 3 weeks to analyze the data collected.

## IV. TESTING HONEYPOT SYSTEM FUNCTIONALITY

In this section, we will test some of the functionalities of the honeypot by running a Nmap scan and brute force attack against the honeypot system. We will target Cowrie, the SSH honeypot container.

*A. Port Scanning the Honeypot System*

I used a Kali Linux Virtual Machine to simulate an attack on the honeypot system. Information on how to install Kali Linux can be found at https://www.kali.org/docs/. Kali Linux comes with Nmap, a port scanning tool, preinstalled on the system. To use this tool, I opened a terminal session and ran the command "nmap -p 22 140.82.3.147" as seen in Figure 8.

```
┌──(kali㉿kali)-[~]
└─$ nmap -p 22 140.82.3.147
Starting Nmap 7.92 ( https://nmap.org ) at 2022-03-05 16:52 EST
Nmap scan report for 140.82.3.147.vultr.com (140.82.3.147)
Host is up (0.020s latency).

PORT   STATE SERVICE
22/tcp open  ssh

Nmap done: 1 IP address (1 host up) scanned in 0.09 seconds

┌──(kali㉿kali)-[~]
└─$
```

Figure 8: Nmap scan ran against the Honeypot System.

This command runs a port scan only checking to see if port 22 is open on the IP address specified. From the results returned from the Nmap scan, we can see that port 22 is open to be used by the SSH Service. This is because as mentioned earlier, Cowrie is a high interaction SSH honeypot designed to log brute force attacks and the shell interaction performed by the attacker once they gain entry.

*A. Brute-Forcing the System*

For simplicity of the experiment and to test the functionality of the honeypot system, I logged into the honeypot SSH server and made the root account accessible over SSH. Also, I changed the password to 12345. I then went to my Kali Linux terminal and used the hydra tool to crack the password for the root account to the honeypot. I ran the command "hydra -l root -P /usr/share/wordlists/Metasploit/unix_password.txt -T 6 ssh://140.82.3.147" as seen below in Figure 9.

Figure 9: Running the Hydra Brute-Force Attack.

The -l option in the command tells hydra to try to log in with the root user. The -P option tells hydra to use the password list specified in the command and the -t option tells hydra how many threads to use. So, in this attack scenario, we used 6 threads. Also, we told it to attack the SSH service on the IP specified. As we can see in Figure 9, we got the exact results that we expected. The password that was cracked for the root user was 12345. Since the test has concluded, I logged back into the SSH server and made it not possible to login to the root user through SSH like how it was prior to me modifying it. This is because this is a good security practice and if it was possible to login to the root user over SSH it would make the attacker suspicious of the honeypot. I then changed the password back to what it was prior. The original password was not that difficult to crack either, but more difficult to crack than 12345. This is because we do still want the attacker to be able to gain access so we can gain more information on what they are looking to do once they are inside the server. I will now let the honeypot system run for a few weeks so it can collect a large amount of data to come back and analyze.

### V. ANALYZING DATA COLLECTED FROM ATTACKS ON THE SYSTEM

In this section, we will analyze the data collected by the honeypot system in the Kibana Dashboard. We will look at the different categories of information collected to see what information we can learn about the attackers. For the most part, we will look at the aggregated data collected by all the containers, but in some cases, we will look at data collected by a specific container.

#### A. Accessing the Data

To access the data, we need to first go back to the T-Pot Landing Page. This can be reached by entering "https://<my.ip>:64297" into a web browser and logging in with the credentials created. I then clicked on "Kibana" to bring me to the Kibana app dashboards page.

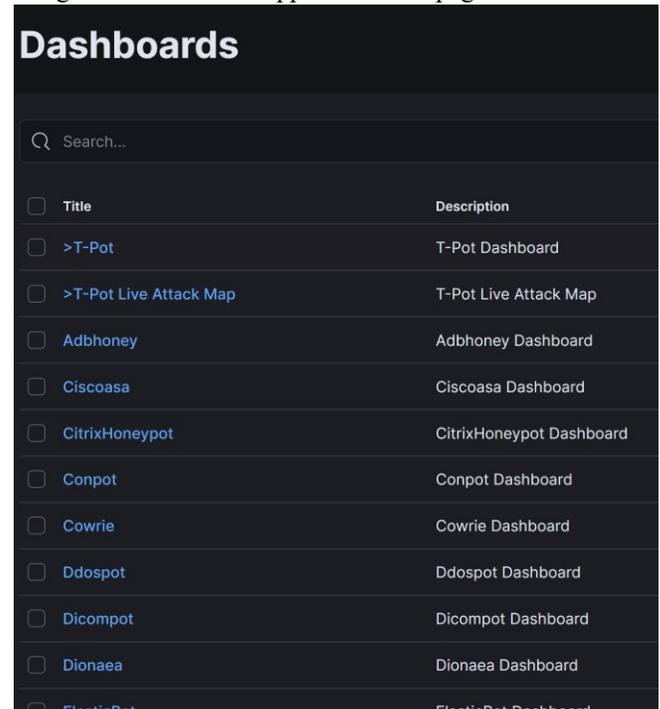

Figure 10: Kibana App Main Dashboards Page.

As seen in Figure 10, a page opens that displays a list of the different honeypot containers deployed. You can reach an individual dashboard by just clicking on the name of that honeypot. You will also notice that the first two titles are
"T-Pot" and "T-Pot Live Attack Map". T-Pot Live Attack Map just shows a global map of where attacks within the T- Pot Network are occurring in the current time. This is not very useful to us. However, the T-Pot Dashboard is the dashboard we will be mainly using because it displays the visuals and data collected by all the honeypots aggregated into one dashboard. Therefore, we will go ahead and click on "T-Pot" to display this dashboard.

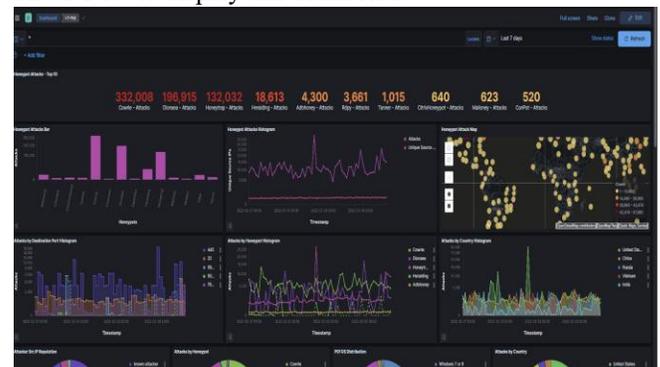

Figure 11: Kibana T-Pot Dashboard.

The dashboard displayed is shown in Figure 11. As mentioned, this dashboard allows us to visualize and view the data collected from all the honeypots in one place. Next, we will be looking at and analyzing some of the charts and data lists that are relevant to gathering intelligence about the attackers.

*B. Most Attacked Honeypots*

At the top of the dashboard, we can see the ten honeypots that experienced the most attacks ranked in sequential order from the most attacked honeypot to the least attacked honeypot.

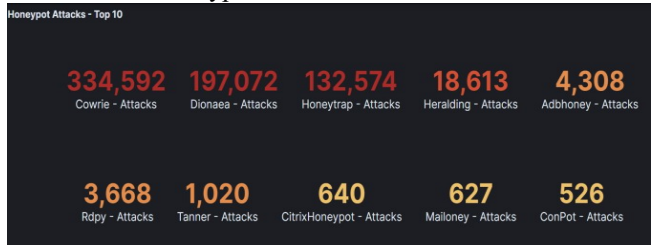

Figure 12: Top 10 attacked honeypots

As we can see in Figure 12, the most attacked honeypot was Cowrie, the SSH honeypot. This is not very surprising since SSH is a very well-understood protocol and can be used to gain remote shell access to a server. If the attacker can abuse the SSH service and gain access to your server, they will be able to obtain a lot of information and have access to your system. From a defender's point of view, this is important information to understand because since SSH is the most attacked protocol, in a live production environment gaining remote access should not be very easy. On top of utilizing a very secure password and SSH keys, MFA should also be configured to gain remote shell access. We can also see that the Dionaea is the second most attacked honeypot, which uses the FTP service to capture attack payloads and malware. This tells us that any way that an attacker can either gain remote access to a server (SSH) or be able to upload files to a server remotely (FTP) is going to attract attackers. This is because the concept of being able to remotely impact a server can have high consequences if abused. As a defender, we can learn from this to make sure that all servers that have remote protocol ports open must be highly secure.

*C. Attacks by Country*

The next graphic we will examine is the countries where the most attacks are from. I will also discuss why that information may and may not be useful.

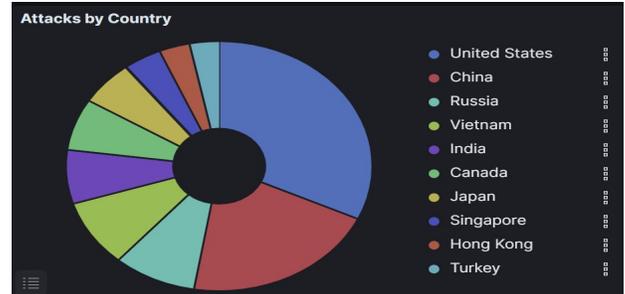

Figure 13: Attacks by Country Pie Chart.

The legend on the right of Figure 13 shows the list of where the most attacks came from in descending order. As we can see, the top 3 countries where the most attacks came from are the United States, China, and Russia. These countries tend to have more sophisticated cyber programs and larger populations, so it is not very surprising. This information can be useful to us because if a specific threat actor is targeting an individual company, we may be able to track where that threat actor is located. However, any decent attacker is most likely using a VPS where they can choose the location of where their IP address is located, or they are using a VPN to hide their true location. That is why focusing on this graphic is not very useful, but it is more important to focus on the IP addresses themselves. We will look at this next.

*D. Attacker Source IP Addresses and their ASNs*

Looking at the Source IP Addresses of where most of the attacks are coming from is very useful to cyber defenders. This is because most attackers have a finite number of resources to work with, so their IP Space is not unlimited.

| Attacker Source IP - Top 10 | |
| --- | --- |
| Source IP | Count |
| 165.22.234.121 | 26,552 |
| 2.56.56.14 | 18,204 |
| 69.171.13.237 | 11,027 |
| 140.82.156.72 | 9,094 |
| 85.100.124.175 | 3,158 |
| 109.94.179.81 | 3,154 |
| 110.227.249.142 | 3,152 |
| 190.75.220.137 | 3,151 |
| 83.52.23.252 | 3,151 |
| 103.43.77.175 | 3,149 |

Figure 14: Top 10 Attacker Source IP Addresses.

In Figure 14, we can see the IP addresses where most of the attacks came from. This information is very useful to cyber defenders because these IP addresses can be blacklisted from the production network. This will not allow these attackers to use these IP addresses on the actual network. The attackers will now have to use other IP addresses to gain access to your resources, therefore making them exhaust their resources. If you continuously blacklist the IP addresses where a lot of the attacks are coming from you can deter attacks from continuing to attack your network. This will help prevent a lot of the noisier attackers from being able to exploit your systems, but you also must be aware that more stealthy attackers will be able to remain under the radar.

From the attacker's IP address, we can figure out the Autonomous System Number (ASN) to learn what organizations are allocating the attacker's IP address.

![Attacker AS/N - Top 10 table]

Figure 15: Top 10 ASNs used by attackers attacking the system.

In Figure 15 we can see that the most popular ASN organization used to attack our honeypot system is Digital Ocean, one of the most popular VPS providers. This shows that a lot of people are abusing the benefits they offer to launch cyber-attacks. There also does appear to be a good number of China-based organizations on the list, helping to support that there might be a lot of attacks launched on the honeypot from China. This information can be useful to us because when we see a user utilizing an IP allocated from Digital Ocean, for example, we can be more cautious and possibly even raise an alert since we know it is commonly used by attackers to attack our systems.

*E. Most Used OS by Attackers*

From our honeypot system, we can gather information about the most popular Operating Systems used by attackers to attack the honeypot system. Kibana displays a pie chart of the operating systems and displays the legend in descending order from most to least used OS.

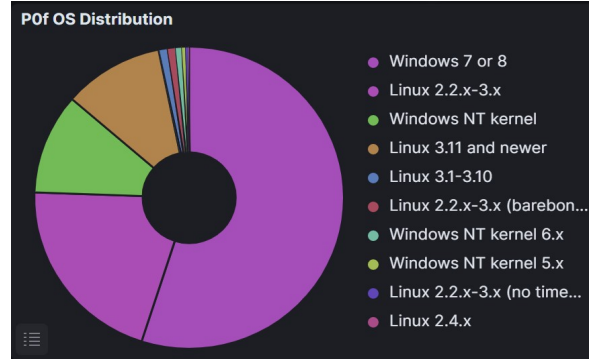

Figure 16: Pie Chart displaying Attacker OS Distributions.

From this pie chart, we can see that Windows 7/8 was the most popular used OS by attackers followed by Linux 2.2.x-3.x. We do not get an exact version number for the most part, but still, get a good idea of the OS the attacker is using. I found it particularly unusual that attackers are running older versions of Windows and Linux, but it can be valuable information. Most common users run Windows 10 as of now, so users running Windows 7/8 can be something to look out for when investigating attackers. A lot of basic users do not run Linux, so anytime there is suspicious activity coming from a user running Linux it can be a red flag.

*F. Most Common Credentials*

In this section, we will look at the most common credentials used by attackers to try to brute-force login to our systems. Many of these usernames and passwords come from dictionary lists of commonly used passwords.

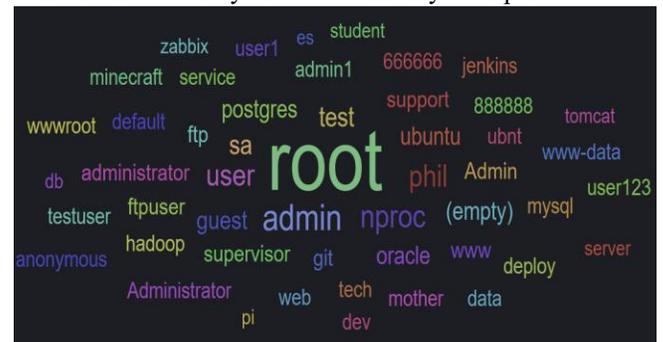

Figure 17: Most Common Usernames attempted by attackers.

In Figure 17, we see the most common usernames attackers tried to use to log in to our honeypot system. As we can see, root and admin were among the most popular. This makes sense because root and admin accounts tend to have higher privileges. This shows us from a defensive perspective why it is so important to disable remote root login because it can be very dangerous if attackers manage to log in.

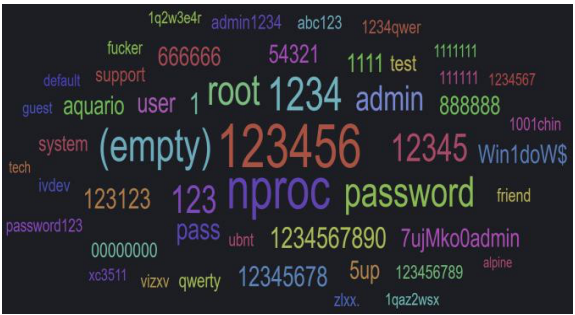

Figure 18: Most Common Passwords attempted by attackers.

In Figure 18, we see the most common passwords attackers tried to use to log in to our honeypot system. Many of these are weak passwords that a basic user might make or default passwords. We can learn from this that it is extremely important to change default passwords and to replace them with a very strong password to prevent attackers from gaining easy access. Passwords should be relatively long and utilize a combination of letters, numbers, and special characters.

*G. Analyzing Suricata Alerts*

In this section, we will analyze the alert raised by Suricata, an open-source IDS/IPS that feeds data into our Kibana dashboard.

| Description | Count |
| --- | --- |
| SURICATA STREAM reassembly sequence GAP -- missing packet(s) | 65,799 |
| ET POLICY SSH session in progress on Expected Port | 13,043 |
| SURICATA STREAM Packet with broken ack | 12,053 |
| SURICATA TCPv4 invalid checksum | 5,008 |
| SURICATA STREAM Packet with invalid timestamp | 3,609 |
| SURICATA Applayer Detect protocol only one direction | 1,553 |
| SURICATA STREAM FIN recv but no session | 1,034 |
| ET SCAN Zmap User-Agent (Inbound) | 998 |
| SURICATA STREAM RST recv but no session | 831 |
| ET INFO Potentially unsafe SMBv1 protocol in use | 740 |

Figure 19: Suricata Alerts.

In Figure 19, we can see the 10 most common alerts detected by Suricata. From this, we can tell that a lot of attackers try to abuse TCP and manipulate the Three-Way TCP Handshake. We can also tell that a lot of users were using a port scanning tool like Nmap because there were a lot of detections of incomplete connections. Finally, we also detected some alerts that SSH and SMB sessions were established. This means that some attackers did gain entry inside our honeypot system, as we expected.

| CVE ID | Count |
| --- | --- |
| CVE-2019-12263... | 62 |
| CVE-2019-0708 ... | 14 |
| CVE-2020-11910 | 6 |

Figure 20: Suricata CVEs Detected.

In Figure 20, we can see the top 3 CVEs detected by Suricata. As mentioned before, a CVE is a common vulnerability and exposure that is known by the public and is detected by a unique signature. CVE-2019-12263 is a high vulnerability with a CVSS score of 8.1. This vulnerability refers to a Buffer Overflow in the TCP component of Wind River VxWorks 6.9.4 and vx7 [16]. The next most common CVE detected is CVE-2019-0708 which has a CVSS score of 9.8, also very high. This vulnerability is a "Remote Desktop Services Remote Code Execution Vulnerability." This means that an attacker can remotely execute code on another user's system [17]. Finally, the third most popular CVE is CVE-2020-11910 which has a CVSS score of 5.3, a medium level severity. This vulnerability is caused by the Treck TCP/IP stack before version 6.0.1.66 having an ICMPv4 Out-of-bounds Read [18]. This means that the system is reading packets that are not in bounds. From all these CVEs detected, it can teach us to make sure that our systems are patched against these popular vulnerabilities. It also shows us how important it is to regularly patch our systems when there are security updates because attackers will try to exploit our systems if they are not patched.

*H. Cowrie Top Shell Commands Executed*

In this section, we will look at the top commands executed inside of the Cowrie honeypot. To look at this, we will go back to the main T-Pot Dashboard page and select "Cowrie Dashboard" to only see the Dashboard for this specific honeypot. As mentioned earlier, Cowrie is a high interaction SSH honeypot that tracks commands entered by attackers into the command line and that data gets sent into Kibana to analyze.

| Command Line Input | Count |
| --- | --- |
| uname -a | 126 |
| cat /proc/cpuinfo \| grep model \| grep name \| ... | 117 |
| cat /proc/cpuinfo \| grep name \| head -n 1 \| a... | 117 |
| cat /proc/cpuinfo \| grep name \| wc -l | 117 |
| crontab -l | 117 |
| free -m \| grep Mem \| awk '{print $2 ,$3, $4, $... | 117 |
| ls -lh $(which ls) | 117 |
| top | 117 |
| uname | 117 |
| uname -m | 117 |

Figure 20: Top 10 Commands Executed by attackers.

In Figure 20, we can see the top 10 commands run by

attackers inside the honeypot. As we can see, the most run command is the "uname" command with the -a flag. We also see many different variations of this command in the top 10 list. This command reveals a lot about the system such as what Linux distro is being run on the system and the version of the distro. This is useful to attackers because if the system is out of date there may be vulnerabilities that the attacker can exploit. Therefore, it is important to update and patch your systems. The next most popular command that we see being run is the attacker looking for information about the system's CPU. An attacker may be interested in the CPU info to see how much processing power your system has. The reason they care about this is that crypto mining is currently extremely popular but requires many resources to drive a profit. So, if attackers can create a botnet of devices that mines for crypto in the background of victims' computers, it will have no cost to the attackers, and they will receive all the benefits.

I. CONCLUSION

A honeypot system can be very beneficial to an organization, but it must be properly implemented and deployed. Honeypots need to be deceptive enough to trick the attacker into thinking that they are in a real system. This will cause the attackers to use up time and resources when attacking the honeypot system, but it will also reveal information about how the attackers are penetrating the system and what they are looking for once they gain access to the system. This information can be useful to the defenders because we can learn how attackers operate and can make sure our actual systems are secure against the various attacks that are being used by attackers. We can also make sure that the information attackers are looking to seek is properly protected.

For future work, we will extend our current cybersecurity framework [19-53] by integrating the Honeypot system as a Cyber Deception Tactic to deceive the attacker.

REFERENCS

[1] B. Lutkevich, "How to build a honeypot to increase network security," WhatIs.com, 31-Mar-2021. [Online]. Available: https://whatis.techtarget.com/feature/How-to-build-a-honeypot-to-increase-network-security. [Accessed: 10-Mar-2022].
[2] J. Xi, "A Design and Implement of IPS Based on Snort," 2011 Seventh International Conference on Computational Intelligence and Security, 2011, pp. 771-773, doithat: 10.1109/CIS.2011.175.
[3] R. Bhardwaj, "Understanding types and benefits of honeypot in network security," IP With Ease, 22-May-2020. [Online]. Available: https://ipwithease.com/understanding-types-and-benefits-of-honeypot-in-network-security/. [Accessed: 10-Mar-2022].
[4] R. Chandel, "Comprehensive guide on honeypots," Hacking Articles, 13-Jan-2022. [Online]. Available: https://www.hackingarticles.in/comprehensive-guide-on-honeypots/. [Accessed: 10-Mar-2022].
[5] "What is a honeypot? how it increases security," Rapid7. [Online]. Available: https://www.rapid7.com/fundamentals/honeypots/. [Accessed: 10-Mar-2022].
[6] Taylor, Jamie, Joseph Devlin, and Kevin Curran. "Bringing location to IP Addresses with IP Geolocation." Journal of Emerging Technologies in Web Intelligence 4.3 (2012).
[7] P. Engebretson and D. Kennedy, The basics of hacking and penetration testing ethical hacking and penetration testing Made Easy. Amsterdam: Syngress, an imprint of Elsevier, 2013.
[8] S. R. Northcutt, "Improve detection using Honeycreds," Improve Detection using HoneyCreds, 01-Jan-1970. [Online]. Available: https://securitywa.blogspot.com/2016/04/improve-detection-using- honeycreds.html. [Accessed: 10-Mar-2022].
[9] "Using honey credentials to make pivoting detectable," LogRhythm, 26-Mar-2020. [Online]. Available: https://logrhythm.com/blog/using-honeywords-to-make-password-cracking-detectable/. [Accessed: 10-Mar- 2022].
[10] N. C. Rowe, " Honeypot Deception Tactics," U.S. Naval Postgraduate School. [Online]. Available: http://faculty.nps.edu/ncrowe/honeypot_deception_tactics.htm. [Accessed: 10-Mar-2022].
[11] Telekom-Security, "Telekom-security/TPOTCE: T-pot - the all in one honeypot platform ," GitHub. [Online]. Available: https://github.com/telekom-security/tpotce. [Accessed: 10-Mar-2022].
[12] Cowrie, "Cowrie/Cowrie: Cowrie SSH/telnet honeypot https://cowrie.readthedocs.io," GitHub. [Online]. Available: https://github.com/cowrie/cowrie. [Accessed: 10-Mar-2022].
[13] T. Werner, "Armedpot/Honeytrap," Honeytrap. [Online]. Available: https://github.com/armedpot/honeytrap/. [Accessed: 10-Mar-2022].
[14] Kibana, "Kibana Guide," Elastic. [Online]. Available: https://www.elastic.co/guide/en/kibana/current/index.html. [Accessed: 10- Mar-2022].
[15] Suricata, "Documentation - Suricata," Suricata, 02-Jun-2021. [Online]. Available: https://suricata.io/documentation/. [Accessed: 10-Mar-2022].
[16] NIST, "CVE-2019-12263 Detail," National Vulnerability Database, 09-Aug-2019. [Online]. Available: https://nvd.nist.gov/vuln/detail/CVE- 2019-12263. [Accessed: 10-Mar-2022].
[17] "CVE-2019-0708 Detail," National Vulnerability Database, 16-May- 2019. [Online]. Available: https://nvd.nist.gov/vuln/detail/CVE-2019-0708. [Accessed: 10-Mar-2022].
[18] "CVE-2020-11910 Detail," National Vulnerability Database, 17-Jun- 2020. [Online]. Available: https://nvd.nist.gov/vuln/detail/CVE-2020- 11910. [Accessed: 10-Mar-2022].
[19] Hisham A. Kholidy, "Multi-Layer Attack Graph Analysis in the 5G Edge Network Using a Dynamic Hexagonal Fuzzy Method",. Sensors 2022, 22, 9.
[20] Hisham A. Kholidy, "Multi-Layer Attack Graph Analysis in the 5G Edge Network Using a Dynamic Hexagonal Fuzzy Method", Sensor Journal. Sensors 2022, 22, 9. https://doi.org/10.3390/s22010009.
[21] Hisham A. Kholidy, Andrew Karam, James L. Sidoran, Mohammad A. Rahman, "5G Core Security in Edge Networks: A Vulnerability Assessment Approach", the 26th IEEE Symposium on Computers and Communications (IEEE ISCC 2021), Athens, Greece, September 5-8, 2021. https://ieeexplore.ieee.org/document/9631531
[22] Hisham A. Kholidy, "A Triangular Fuzzy based Multicriteria Decision Making Approach for Assessing Security Risks in 5G Networks", December 2021, Preprint={2112.13072}, arXiv.
[23] Kholidy, H.A., Fabrizio Baiardi, "CIDS: A framework for Intrusion Detection in Cloud Systems", in the 9th Int. Conf. on Information Technology: New Generations ITNG 2012, April 16-18, Las Vegas, Nevada, USA. http://www.di.unipi.it/~hkholidy/projects/cids/
[24] Kholidy, H.A. (2020), "Autonomous mitigation of cyber risks in the Cyber–Physical Systems", doi:10.1016/j.future.2020.09.002, Future Generation Computer Systems,Volume 115, 2021, Pages 171-187, ISSN 0167-739X, https://doi.org/10.1016/j.future.2020.09.002.
[25] Hisham A. Kholidy, Abdelkarim Erradi, Sherif Abdelwahed,

Fabrizio Baiardi, "A risk mitigation approach for autonomous cloud intrusion response system", Computing Journal, Springer, DOI: 10.1007/s00607-016-0495-8, June 2016. (Impact factor: 2.220). https://link.springer.com/article/10.1007/s00607-016-0495-8

[26] Hisham A. Kholidy, "Detecting impersonation attacks in cloud computing environments using a centric user profiling approach", Future Generation Computer Systems, Vol 115, 17, December 13, 2020, ISSN 0167-739X.

[27] Kholidy, H.A., Baiardi, F., Hariri, S., et al.: "A hierarchical cloud intrusion detection system: design and evaluation", Int. J. Cloud Comput., Serv. Archit. (IJCCSA), 2012, 2, pp. 1–24.

[28] Kholidy, H.A., "Detecting impersonation attacks in cloud computing environments using a centric user profiling approach", Future Generation Computer Systems, Volume 115, issue 17, December 13, 2020, Pages 171-187, ISSN 0167-739X, https://doi.org/10.1016/j.future.2020.12.

[29] Kholidy, Hisham A.: 'Correlation-based sequence alignment models for detecting masquerades in cloud computing', IET Information Security, 2020, 14, (1), p.39-50, DOI: 10.1049/iet-ifs.2019.0409.

[30] Kholidy, H.A., Abdelkarim Erradi, "A Cost-Aware Model for Risk Mitigation in Cloud Computing SystemsSuccessful accepted in 12th ACS/IEEE International Conference on Computer Systems and Applications (AICCSA), Marrakech, Morocco, November, 2015.

[31] A H M Jakaria, Mohammad A. Rahman, Alvi A. Khalil, Hisham A. Kholidy, Matthew Anderson et al "Trajectory Synthesis for a UAV Swarm Based on Resilient Data Collection Objectives", IEEE Transactions on Network and Service Management, November, 2022 doi: 10.1109/TNSM.2022.3216804.

[32] Hisham A. Kholidy, Andrew Karam, James Sidoran, et al. "Toward Zero Trust Security in 5G Open Architecture Network Slices", the 40th IEEE Military Conference (MILCOM), San Diego, CA, USA, November 29, 2022.

[33] Hisham A. Kholidy, Riaad Kamaludeen "An Innovative Hashgraph-based Federated Learning Approach for Multi Domain 5G Network Protection", IEEE Future Networks (5G World Forum), Montreal, Canada, October 2022.

[34] Hisham A. Kholidy, Andrew Karam, Jeffrey H. Reed, Yusuf Elazzazi, "An Experimental 5G Testbed for Secure Network Slicing Evaluation", IEEE Future Networks (5G World Forum), Montreal, Canada, October 2022.

[35] Hisham A. Kholidy, Salim Hariri, Pratik Satam, Safwan Ahmed Almadani "Toward an Experimental Federated 6G Testbed: A Federated learning Approach", the 13th Int. Conf. on Information and Communication Technology Convergence (ICTC), Jeju Island, Korea, October 9, 2022.

[36] NI Haque, MA Rahman, D Chen, Hisham Kholidy, "BIoTA: Control-Aware Attack Analytics for Building Internet of Things", 2021 18th Annual IEEE International Conference on Sensing, Communication

[37] Kholidy, H.A., Ali T., Stefano I., et al, "Attacks Detection in SCADA Systems Using an Improved Non-Nested Generalized Exemplars Algorithm", the 12th IEEE International Conference on Computer Engineering and Systems (ICCES 2017), December 19-20, 2017.

[38] Qian Chen, Kholidy, H.A., Sherif Abdelwahed, John Hamilton, "Towards Realizing a Distributed Event and Intrusion Detection System", the Int. Conf. on Future Network Systems and Security, Florida, USA, Aug 2017.

[39] Hisham A. Kholidy, Abdelkarim Erradi, Sherif Abdelwahed, Abdulrahman Azab, "A Finite State Hidden Markov Model for Predicting Multistage Attacks in Cloud Systems", in the 12th IEEE Int. Conf. on Dependable, Autonomic and Secure Computing, China, August 2014.

[40] Ferrucci, R., & Kholidy, H. A. (2020, May). A Wireless Intrusion Detection for the Next Generation (5G) Networks", Master's Thesis, SUNY poly.

[41] Rahman, A., Mahmud, M., Iqbal, T., Saraireh, L., Hisham A. Kholidy., et. al. (2022). Network anomaly detection in 5G networks. Mathematical Modelling of Engineering Problems, Vol. 9, No. 2, pp. 397-404. https://doi.org/10.18280/mmep.090213

[42] Hisham Kholidy, "State Compression and Quantitative Assessment Model for Assessing Security Risks in the Oil and Gas Transmission Systems", doi : 10.48550/ARXIV.2112.14137, https://arxiv.org/abs/2112.14137}, December 2021.

[43] Hisham A. Kholidy, "Correlation Based Sequence Alignment Models For Detecting Masquerades in Cloud Computing", IET Information Security Journal, DOI: 10.1049/iet-ifs.2019.0409, Sept. 2019 (ISI Impact Factor(IF): 1.51) https://digital-library.theiet.org/content/journals/10.1049/iet-ifs.2019.0409

[44] Hisham A. Kholidy, "An Intelligent Swarm based Prediction Approach for Predicting Cloud Computing User Resource Needs", the Computer Communications Journal, December 19 (ISI IF: 2.766). https://www.sciencedirect.com/science/article/abs/pii/S0140366419303329

[45] Hisham A. Kholidy, Abdelkarim Erradi, "VHDRA: A Vertical and Horizontal Dataset Reduction Approach for Cyber-Physical Power-Aware Intrusion Detection Systems", SECURITY AND COMMUNICATION NETWORKS Journal (ISI IF: 1.376), March 7, 2019. vol. 2019, Article ID 6816943, 15 pages. https://doi.org/10.1155/2019/6816943.

[46] Hisham A. Kholidy, Hala Hassan, Amany Sarhan, Abdelkarim Erradi, Sherif Abdelwahed, "QoS Optimization for Cloud Service Composition Based on Economic Model", Book Chapter in the Internet of Things. User-Centric IoT, Volume 150 of the series Lecture Notes of the Institute for Computer Sciences, Social Informatics and Telecommunications Engineering pp 355-366, June 2015.

[47] Hisham A. Kholidy, Alghathbar Khaled s., "Adapting and accelerating the Stream Cipher Algorithm RC4 using Ultra Gridsec and HIMAN and use it to secure HIMAN Data", Journal of Information Assurance and Security (JIAS), vol. 4 (2009)/ issue 4, pp 274-283, 2009. (Indexed by INSPEC, Scopus, Pubzone, Computer Information System Abstracts, MathSci).

[48] Hisham A. Kholidy, "Towards A Scalable Symmetric Key Cryptographic Scheme: Performance Evaluation and Security Analysis", IEEE International Conference on Computer Applications & Information Security (ICCAIS), Riyadh, Saudi Arabia, May 1-3, 2019. https://ieeexplore.ieee.org/document/8769482

[49] Samar SH. Haytamy, Hisham A. Kholidy, Fatma A. "ICSD: Integrated Cloud Services Dataset", Springer, Lecture Note in Computer Science, ISBN 978-3-319-94471-5, https://doi.org/10.1007/978-3-319-94472-2.

[50] Stefano Iannucci, Hisham A. Kholidy Amrita Dhakar Ghimire, Rui Jia, Sherif Abdelwahed, Ioana Banicescu, "A Comparison of Graph-Based Synthetic Data Generators for Benchmarking Next-Generation Intrusion Detection Systems", IEEE Cluster 2017, Sept 5 2017, Hawaii, USA.

[51] Mustafa, F.M., Kholidy, H.A., Sayed, A.F. et al. Enhanced dispersion reduction using apodized uniform fiber Bragg grating for optical MTDM transmission systems. Opt Quant Electron 55, 55 (2023). https://doi.org/10.1007/s11082-022-04339-7

[52] Abuzamak, M., & Kholidy, H. (2022). UAV Based 5G Network: A Practical Survey Study. arXiv. https://doi.org/10.48550/arXiv.2212.13329

[53] Abuzamak, M., & Kholidy, H. (2022). UAV Based 5G Network: A Practical Survey Study. arXiv. https://doi.org/10.48550/arXiv.2212.13329